\documentclass[twocolumn,preprintnumbers,amsmath,amssymb,prl]{revtex4}

\usepackage[dvips]{graphicx}
\usepackage{dcolumn}
\usepackage{bm}
\usepackage[dvips]{epsfig}

\begin{document}

\title{Acoustic emission associated with the bursting of a gas bubble\\ at the free surface of a non-newtonian fluid.}

\author{T. Divoux, V. Vidal$\dag$, F. Melo and J.-C. G\'eminard$\dag$}
\affiliation{Departamento de F\'{\i}sica, and Center for Advanced Interdisciplinary
Research in Materials (CIMAT),  Universidad de Santiago de Chile (USACH),
Av. Ecuador 3493, Casilla 307, Correo 2, Santiago de Chile, Chile.\\
$\dag$Permanent address : Universit\'e de Lyon,
Laboratoire de Physique, Ecole Normale Sup\'erieure de Lyon - CNRS,
46, All\'ee d'Italie, 69364 Lyon cedex 07, France.}

\begin{abstract}
We report experimental measurements of the acoustic emission
associated with the bursting of a gas bubble at the free surface of
a non-newtonian fluid. On account of the viscoelastic properties of the
fluid, the bubble is generally elongated. The associated frequency
and duration of the acoustic signal are discussed with regard to the
shape of the bubble and successfully accounted for by a simple linear
model. The acoustic energy exhibits a high sensitivity to
the dynamics of the thin film bursting, which demonstrates that, in
practice, it is barely possible to deduce from the acoustic
measurements the total amount of energy released by the event. 
Our experimental findings provide clues for the understanding of the
signals from either volcanoes or foams, where one observes
respectively, the bursting of giant bubbles at the free surface of
lava and bubble bursting avalanches.
\end{abstract}

\maketitle

\section{Introduction}

A host of broad-interest phenomena involve bursting bubbles at fluid
surfaces. In daily life, jam or pur\'ee cooking produces sonic
bubbles that can project fragments at bursting. In the geophysical
context, giant bubbles bursting at the top of a volcano vent, or at
the surface of a lava lake, are examples whose understanding might
be crucial for predicting volcanic activity \cite{parfitt04,gonnermann03}.
 Although less considered, bubble bursting also occurs at the surface of aqueous
 foams \cite{weaire99}
 typically produced by wash- or beauty- products or even by poured beer.
 The analysis of acoustics emission is then a natural way of investigating bursting
systems, revealing the collapse or bursting mechanism and properties of
the fluid gas mixture \cite{prosperetti93,leighton94,muller99,
vandewalle01,vandewalle02,vandewalle03}. 

On the one hand, various statistical analysis of bursting noise have been carried out 
\cite{muller99,vandewalle01,ding07}. For instance, the sound pattern of
collapsing foams was recently analyzed, revealing a log-normal
distribution for the energy of events. For the events of highest
acoustic energy, the distribution is however a power law suggesting
that a "wide variety of bubble membranes areas is exploding"
\cite{vandewalle01}. Consistently, the film rupture event seems to
be independent of the bubble size and exhibits instead a correlation in
space due to cascade bursting \cite{vandewalle03}. In turn, the
typical frequency of the acoustic signal has been
statistically correlated to the bubble size \cite{vandewalle01,ding07}.
The acoustic emission of a single standing spherical bubble has been
recorded as well \cite{ding07} and sophisticated high speed
techniques have been used to elucidate the bursting dynamics of
spherical smectic films \cite{muller07}. However, detailed
correlations of high speed images of foam bursting-films with the
features of the acoustic emission have not yet been performed.

On the other hand, volcanologists have recorded the sound produced
by astounding burstings, and have tried to infer, from the signal
characteristics, the dynamical processes involved in these natural
phenomena \cite{Johnson03}. Previous laboratory experiments have intended to
reproduce bubbles formation, rising and bursting in geometries
presenting similarities with bubbles in magma conduit or lava lakes
\cite{jaupart,Ripepe01}. Since the forces applied on lava have a
time scale much larger than its relaxation time (about $4 \times
10^{-8}$~s for Strombolian magma \cite{Webb90}), most of model
experiments have been performed in newtonian fluid
\cite{jaupart,Vergniolle96a,Ripepe01}. However, departure from this
newtonian behavior can occur if the magma contains crystals
\cite{Vergniolle96a}, which can be observed for example on
Strombolian ejecta \cite{Francalanci89}. Some experiments, performed
in more 'exotic' fluids like japanese curry or tomato sauce, have
demonstrated the complex behavior of such fluids, and brought to the fore the
non-trivial physical processes leading to sound generation - from
Helmholz resonator type to bubble oscillation inside the fluid \cite{Ichihara05}.

\begin{figure*}[!t]
\begin{center}
\includegraphics[width=1.6\columnwidth]{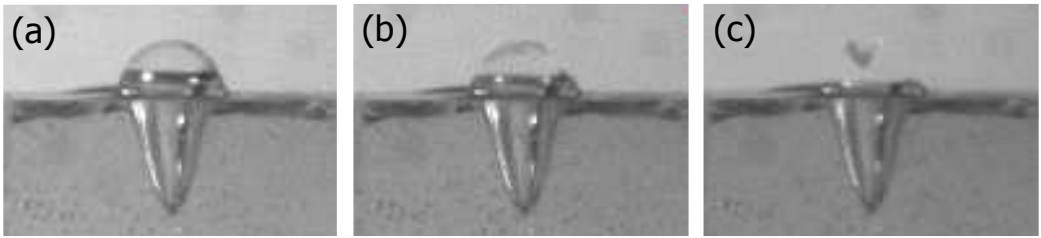}
\end{center}
\caption{\small{{\bf Bursting event in the fast camera.}
    (a) Initially, the bubble sits at the free surface. The film at the top is thinning due to the drainage.
    (b) The thin film suddenly breaks. One observes, in this specific example, that the film tears at its base.
    (c) As a result of the outgoing air flow and of the capillary forces, the remaining part of the bubble head
    is blown upwards and shrinks. The width of each image is 3~cm whereas the time-difference between them is 0.8~ms.
    Here, the acoustic emission associated with the event does not last more than 2~ms, during which the
    bubble body clearly does not significantly deform.}}
\label{fig:film}
\end{figure*}

To understand the geometrical and dynamical aspects of sound
produced by bursting bubbles having an elongated shape, we described
recently the acoustic emission from an overpressurized cylindrical
cavity, closed at one end by a fluid film \cite{Vidal06}. We showed
that the cavity geometry governs the frequency, the viscous
dissipation and radiation are responsible for the wave damping and
the acoustic energy depends not only on the energy initially loaded inside the
cavity but also on the characteristic time associated with the
film bursting.

Here, we present the analysis of the high-frequency acoustic-wave
emitted by a bubble bursting at the free surface of a non-newtonian
fluid. In such fluids, the complex rheology \cite{Bird87} is the
source of puzzling phenomena including surface instabilities due to elastic effects
\cite{Podgorski02a}, cusp at the tail of elongated rising bubbles
\cite{Hassager79,Bird87,Liu95,Chhabra} and oscillations of falling
spheres or rising air-bubbles \cite{Belmonte00,Jayaraman03,Handzy04}. 
We choose an experimental situation in which the bubble, generally elongated,
exhibits a nearly-conical steady-shape during the rise toward the free surface.
When the bubble reaches the free surface, a liquid film separates
the bubble body from the surrounding air. This film thins, eventually breaks, 
and can display a complex behavior (see for instance \cite{Debregeas95,Debregeas98,Dalkoni99,Dutcher00}
for viscous films, and \cite{Warszynski96,Angarska01} for soap films).

In the chosen experimental conditions, due to the non-newtonian fluid properties,  
the bubble body does not significantly deform during the bursting event.
We first link the characteristics of the radiated sound with bubble volume 
and shape, which is governed by the rheological properties of the fluid.
Despite the conical shape of the bubbles, acoustic signals show a narrow
frequency spectrum whose characteristic wavelength is linear in the
bubble length, exhibiting a well-defined offset that is due to
acoustic radiation \cite{Levine48}. Both the results are theoretically
explained using linear acoustic \cite{Pierce89,Landau87,Rayleigh40}.
Differences and similarities with the newtonian-fluid case are
underlined. We then show that, if the rupture time of the viscoelastic film 
does not control the wave generation (i.e., it is fast enough),  
an optimal size for a 'sonic' bubble exists: The largest amplitude of the acoustic signal
is recorded for a bubble having this optimal size.

We then show that, without a detailed knowledge of film-bursting dynamics,
acoustic measurements are not a reliable method to access the total amount
of energy released.
Indeed, the amount of the energy transferred to the acoustic wave drastically
depends on the characteristic time associated with the opening of the bubble \cite{Vidal06},
which is not controlled experimentally. 
This result might find interesting
applications to aqueous foams as well, indicating that the statistic
of energy released by bubble bursting-avalanches, recently
characterized by acoustic emission \cite{vandewalle01}, might be not
only influenced by the distribution of the bubble sizes, but also by that 
of the rupture times.
Consistently, our results suggest that bursting cascades 
might be triggered more likely by silent bubbles than by noisy ones.  In this case, most of the
potential energy loaded inside the cavity would contribute to larger
distortions of bubble network.

\section{Experimental setup and procedure}

The experimental setup consists of a vertical plexiglas container
(square section $30 \times 30$~mm, height 88~mm) filled with a
transparent non-newtonian fluid up to the upper plane. Thanks to the
transparent and planar walls of the container, the fluid can be
imaged from the side without any optical distortion by means of a
fast-camera  (HiSIS 2002, KSV Instruments Ltd., up to 1220
images/sec, Fig.~\ref{fig:film})

In order to produce bubbles having a well-defined volume $V$ (from
0.1 to 1.5~mL), a chosen amount of air is rapidly injected by means
of a syringe pump connected to the container by a hole
drilled at the center of the lower plane.
After injection, the bubble rises in the fluid, reaches the upper 
free surface (Fig.~\ref{fig:film} \& \ref{fig:bubble}) and finally 
bursts, producing a characteristic audible sound.

To characterize this phenomenon, we record the acoustic
emission by means of a microphone (ATM33a, Audio Technica associated
with a preamplifier, Eurorack UB802) which is
located $3$~cm away from the gel free-surface, with a 45-degrees
inclination from the vertical. The position of the microphone shall remain
identical for all the experimental results reported herein.

The chosen non-newtonian fluid is obtained by diluting a commercial
hair-dressing gel ({\it Gel fijador de cabello, for men, Camel
White$\tiny~^{\textregistered}$}) in pure water. This latter choice
is mainly justified by the fact that, in such a fluid, the air
bubbles usually exhibit a nice vertically-elongated shape,
terminated by a cusp at the bottom, which significantly differs from
the rounded shape usually observed in a newtonian fluid
\cite{Bird87}. In addition, one can
easily be supplied with large quantities of fluid, reproducible
mixtures are rather easy to prepare and they are stable in the time.
The non-newtonian character of the fluid is more or less pronounced
depending on the concentration, $c$, of gel in the mixture (from 25
to 40\% in volume). All the solutions are obtained after the mixing
of the two components during half a day by means of a magnetic
stirrer. Afterwards, the small bubbles that still remain trapped in
the fluid are eliminated by placing the solution in an ultrasonic bath for several hours.
In order to avoid any memory effect, the gel is stirred and let at rest for a few
minutes between two bubble rises.
The mixtures are likely to be subjected to drying: In order to avoid any significant
change in the overall concentration $c$, each sample is used only for 4 days.

\section{Preliminary observations : Bubble shape and bursting dynamics}

Before analyzing in details the acoustic emission associated with
the bursting of a single bubble at the free surface, one must first
pay attention to the bubble shape and dynamics.

First, depending on the concentration $c$ and on the volume $V$, one
observes two qualitative different steady shapes of the bubbles which rise
up in the bulk of the fluid. Indeed, for low gel concentration
(typically $c < 30\%$), the bubbles exhibit an almost spherical
shape, similar to that observed in newtonian fluids. On the
contrary, for larger concentration (typically $c > 30\%$), the
bubbles exhibit an elongated shape with a cusped tail, as already
observed in non-newtonian fluids \cite{Bird87,Hassager79,Chhabra,Liu95}.
During the rise, one also notices oscillations in the bubble shape,
as previously pointed out in the literature \cite{Belmonte00}.

\begin{figure*}[!t]
\begin{center}
\includegraphics[width=2\columnwidth]{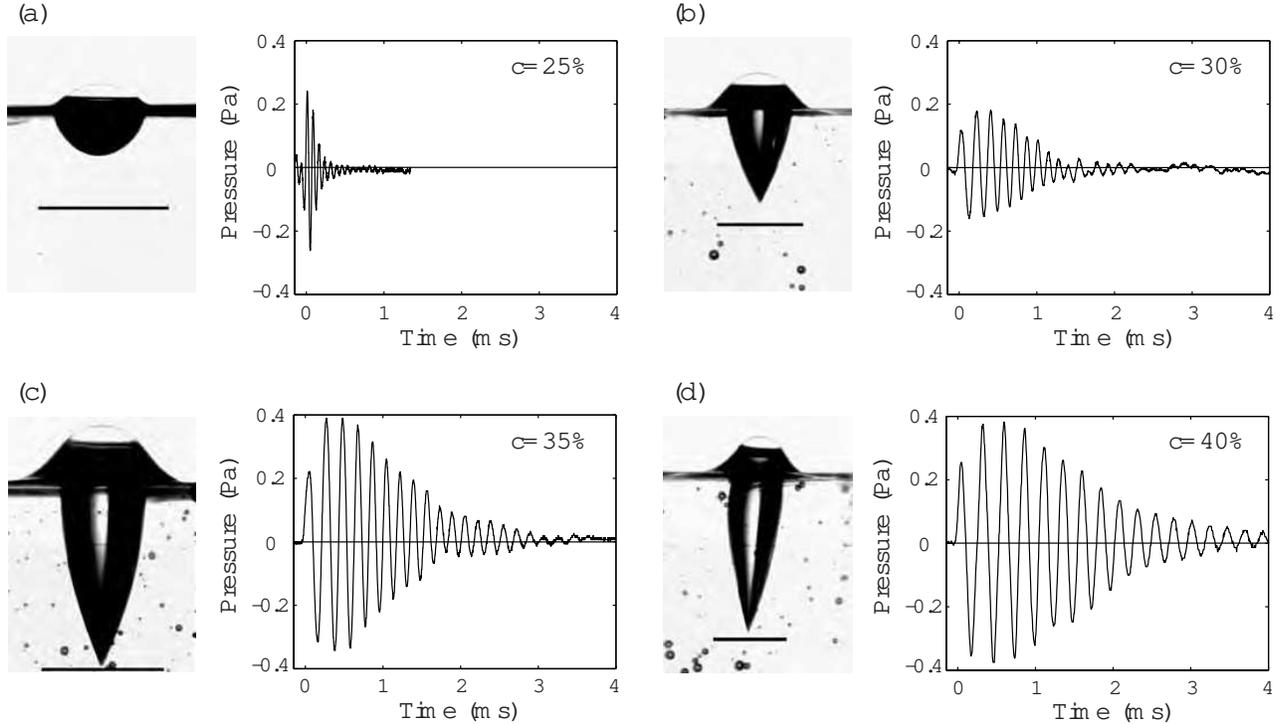}
\end{center}
\caption{\small{{\bf Images of the bubble right before bursting and associated acoustic signals.}
    In the images (scale bar: 1~cm), one notices that the bubbles are more elongated
    when the gel concentration $c$ is larger. In addition, we report the signal
    from the microphone.
    We observe that the typical frequency decreases and the characteristic duration of the acoustic emission
    increases when the bubble length is increased.}}
\label{fig:bubble}
\end{figure*}

The shape of the bubble which bursts at the free surface
qualitatively exhibits almost the same transition. For small
concentration $c$, one observes a rounded bubble on the last image
previous to the bursting (Fig.~\ref{fig:bubble}a) whereas, for a
larger concentration $c$, a cusp is clearly observed
(Fig.~\ref{fig:bubble}d). However, we point out that at intermediate
concentration (typically $c \sim 30\%$) the shape of the bubble,
when bursting, also depends on its dynamical behavior when reaching
the free surface and, thus, on its volume $V$: For small volume ($V
< 0.25$~mL), only rounded bubbles are observed whereas, for large
volume ($V > 0.40$~mL), the bubbles always exhibit a cusp; In the
intermediate range ($0.25 \leq V < 0.40$~mL), both types of bubble
can be observed for the same volume $V$. 
Qualitatively, the capillary forces are large
enough to maintain the small bubbles in equilibrium at the free
surface so that a bubble, which initially exhibits a cusp while
rising, deforms and equilibrates before the thin film that encloses
the inside air breaks due to the drainage. No cusp is then observed.
On the contrary, the capillary forces are not large enough to
maintain the largest  bubbles in equilibrium at the free surface. As
a consequence, the tail of  the bubble does not significantly deform
at the free surface before the thin film at the top breaks, due to the increase
of its surface area and not to the drainage, in this case. A cusp is thus systematically
observed. In the intermediate case, the capillary forces are likely
to maintain the bubble at the free surface but, because of some
premature ruptures of the thin film at the top, one can observe
a bubble exhibiting either a rounded bottom or a cusp.

The equilibrium shape or the dynamical behavior of the bubble at the
free surface, which would deserve an extensive study to be accounted
for, are not the aim of the present work. From the qualitative
description of the bubble behavior at the free surface presented
above, we shall only remember that the bubble are elongated and
exhibit a cusp in most of the experimental conditions and that,
depending on the volume, the thin film at the top might break either
because of the drainage or of the increase in its surface area. In
what follows, we shall only analyze the acoustic signal associated with
the bursting event with regard to the shape and dynamics of the bubble 
at the free surface.

\section{The acoustic signal}

The bursting event is systematically associated with the emission of a sound wave,
characterized by a well-defined frequency (Fig.~\ref{fig:bubble}).

Qualitatively, the sudden bursting of the thin film at the top
excites a resonant pressure-wave in the bubble body which is
initially overpressurized. The phenomenon is similar to that
thoroughly described in Ref.~\cite{Vidal06}, where the acoustic
emission associated with the bursting of a thin soap film that
initially closes a cylindrical overpressurized cavity is analyzed in
details. In our experimental case, due to the viscoelastic
properties of the fluid, the bubble body is generally elongated and
the bubble wall does not significantly deform during the
characteristic duration of the sound emission (Fig.~\ref{fig:film}).
The opened bubble-body thus selects resonant modes among which the
fundamental is the most intense. Due to the radiation at the open
end, one records outside the cavity a sound wave exhibiting
well-defined frequency and duration which are the subject of the
analysis presented below.

\begin{figure}[!t]
\begin{center}
\includegraphics[width=\columnwidth]{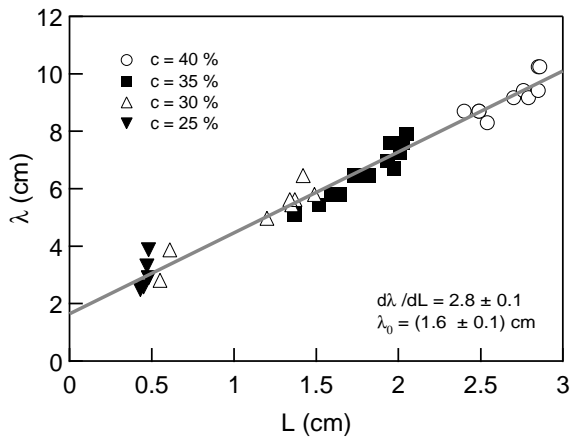}
\end{center}
\caption{\small{{\bf Wavelength $\lambda$ vs. bubble length $L$}.
    The resonant wavelength increases linearly with the bubble length, estimated from
    the open aperture to the tail. The offset $\lambda_0$ is accounted for by the radiation at
    the open-end.}}
\label{fig:lambda}
\end{figure}

\subsection{Acoustic wavelength}

\begin{figure}[!h]
\begin{center}
\includegraphics[width=\columnwidth]{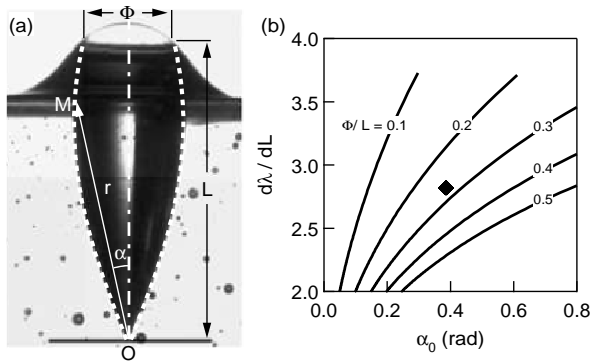}
\end{center}
\caption{\small{{\bf (a) Image of a bubble and definitions.}
    The dotted curve corresponds to the proposed interpolation of the bubble profile by
    $\alpha(r) = \alpha_0 \cos{(\beta \frac{r}{L})}$ with $\alpha_0 = 0.378$ and $\beta = 1.11$.
    {\bf (b) Slope $\frac{d\lambda}{dL}$ vs. $\alpha_0$ for different ratio $\frac{\Phi}{L}$.}
    The slope $\frac{d\lambda}{dL}$ is obtained from the solution of Eq.~\ref{propagation}
    taking into account the profile of the bubble wall. The full diamond corresponds to the
    bubble in (a).}}
\label{fig:sketch}
\end{figure}

In figure \ref{fig:lambda}, we report the wavelength, $\lambda$,
associated with the acoustic wave in air as a function of the bubble
length, $L$, as defined in figure \ref{fig:sketch}. Taking the
whole set of the experimental data obtained for different gel
concentration $c$ and bubble volume $V$ into account, one observes
experimentally that $\lambda$ increases almost linearly with $L$
according to $\lambda \simeq \lambda_0 + (2.8\pm0.1) L$
with $\lambda_0 \simeq (1.6\pm0.1)~$cm. The experimental slope
$d\lambda/dL \simeq 2.8$ deserves to be contrasted with the slope
$d\lambda/dL = 4$ obtained in the case of a cylindrical cavity
\cite{Pierce89,Vidal06}. In the same way, a straightforward analysis of the
acoustic problem leads to $d\lambda/dL = 2$ in the case of a conical
cavity. In order to account for the experimental slope, let us now
consider the acoustic wave inside the bubble, taking into account the
cusp at the tail and the overall shape of the cavity.

Because of the cusp at the tail, the bubble body resembles a cone.
As a consequence, we shall work in a system of spherical coordinates
centered in $O$, at the cusp. Let $M$ be a point of the bubble wall.
The shape of the bubble, assumed to be axisymmetric, can be
accounted for by the angle $\alpha(r)$ between the vector $\overrightarrow{OM}$
and the symmetry axis ($r \equiv OM$,
Fig.~\ref{fig:sketch}a). For instance, a conical bubble would be
described by $\alpha(r) = \alpha_0$, a constant. In a first
approximation, assuming that the variation of the bubble
cross-section does not depend too rapidly (in comparison to the
wavelength) on the radius $r$, we can write the equation for the
pressure field $P(r,t)$, assumed to depend only on the distance $r$,
\begin{eqnarray}
\frac{1}{v^2} \frac{\partial^2 P(r,t)}{\partial t^2}
&&= \frac{1}{r^2}\frac{\partial}{\partial r}\left(r^2 \frac{\partial P(r,t)}{\partial r} \right)\label{propagation}\\
&&+ \frac{\partial P(r,t)}{\partial r} \frac{d}{d r} \Bigl(\log{[1-\cos \alpha(r)]}\Bigr)\nonumber
\end{eqnarray}
where $v$ stands for the velocity of the sound in air.
The equation (\ref{propagation}), written in spherical coordinates,
governs the propagation of a pressure wave in an acoustic horn \cite{Rayleigh40,Pierce89}
whose profile is described by the function $\alpha(r)$.
First, a velocity node locates at $O$. Second, neglecting the radiation at the open end,
we can, in a first approximation, assume that a pressure node locates in the plane of the aperture
[$P(L) = 0$, note here that the condition of zero pressure in the output plane is not compatible
with the geometry of the pressure field. However, in a first approximation, to within a term of the order
of {\bf $\Phi / L$}, where $\Phi$ stands for the aperture diameter, this assumption provides a good estimate of the resonant wavelength.]
For instance, in the case of a conical cavity ($\alpha = \alpha_0$), the solution at the
frequency $\omega$ is $P(r,t) = \frac{\sin{k r}}{k r} e^{j \omega t}$ with $k L = \pi$,
which leads to $\lambda = 2 L$ for the fundamental $\left(k \equiv \frac{2\pi}{\lambda}\right)$.
Experimentally, the bubble is not conical
and its profile is successfully interpolated by the phenomenological function
$\alpha(r) = \alpha_0 \cos{(\beta \frac{r}{L})}$,
where $\alpha_0$ denotes half the angle at the bubble tail (Fig.~\ref{fig:sketch}a).
In this case, the equation (\ref{propagation}) must be solved
numerically.
The coefficient $\beta$ accounts for the diameter $\Phi$ of the aperture
at the top surface, according to $\alpha_0 \cos{(\beta \sqrt{1+({\frac{\Phi}{2L}})^2})} = \arctan \left(\frac{\Phi}{2 L}\right)$.
In Fig.~\ref{fig:sketch}b, we report the slope $\frac{d\lambda}{dL}$ as a function
of $\alpha_0$ for different values of the ratio $\Phi/L$.
For instance, one obtains $\frac{d\lambda}{dL} = 2.8$ for $\alpha_0 = \frac{\pi}{4}$ and $\frac{\Phi}{L} = \frac{1}{2}$,
and $\frac{d\lambda}{dL} = 2.9$ for $\alpha_0 = \frac{\pi}{8}$ and $\frac{\Phi}{L} = \frac{1}{4}$, which corresponds to
the typical values of these two parameters in our experimental conditions.
Thus, in spite of the slight dependance of the slope $\frac{d\lambda}{dL}$ on the $\alpha_0$ and $\Phi/L$,
the experimental wavelength $\lambda$ is observed to depend almost linearly on the bubble length $L$
with $\frac{d\lambda}{dL} \simeq 2.8$ for the whole set of experimental data reported in Fig.~\ref{fig:lambda}.

However, the simple acoustic model presented above fails in accounting for the finite offset
$\lambda_0 \simeq 1.6$~cm clearly observed in figure \ref{fig:lambda}.
As already pointed out, we arbitrarily assumed that a pressure node locates in the plane
of the open end of the bubble, this latter condition being incompatible with the
geometry of the pressure field inside the bubble. In order to recover the value $\lambda_0$
of the offset, one must consider the diffraction of the sound wave by the aperture.
Indeed, the boundary condition in the output plane is imposed by the continuity of the pressure
and velocity fields in this very plane, and one must also consider the acoustic wave outside the cavity.
The structure of the acoustic wave resulting from the diffraction of a planar wave by a
circular aperture (diameter $\Phi$) has been determined by several authors.
As a result, to the first order in $\Phi/\lambda$, one obtains the acoustic impedance of the outer
acoustic wave ${\cal Z}_{out} \simeq \zeta j k \Phi$, where $\zeta=\frac{4}{3\pi}$ for of a flanged aperture
\cite{Rayleigh40} and $\zeta=0.3$ for an unflanged aperture \cite{Levine48}.
In the case of the diffraction of a spherical wave, we expect these results to hold true
[the correction due to the curvature of the pressure field is expected to be of the order of
$\left(\frac{\Phi}{L}\right)^2$.]
In addition, we can estimate the acoustic impedance of the inner acoustic wave, ${\cal Z}_{in}$,
by calculating the average pressure and velocity in the aperture plane from the solution of equation
(\ref{propagation}), the boundary condition at $O$ being taken into account.
Then, insuring the continuity of the pressure and velocity fields in the aperture plane,
one can determine the resonant wavelength $\lambda$.
In the case of a conical cavity, writing ${\cal Z}_{out}/(\rho v) \simeq \zeta j k \Phi$,
one gets $\lambda = 2 L + 2 \pi \zeta \Phi$. Thus, in the case of a conical cavity and of
a flanged aperture, we estimate $\lambda_0 \simeq \frac{8}{3}\Phi$ (We remind here that
one would expect $\lambda_0 \simeq \frac{4}{3\pi}\Phi$ for a cylindrical cavity \cite{Vidal06}.)
In the case of the bubble, the impedance ${\cal Z}_{in}$ must be evaluated numerically as
the equation (\ref{propagation}) does not exhibit any simple analytical solution.
Here, we only aim at elucidating the physical origin of the offset $\lambda_0$.
As a consequence, we only point out that, if due to the radiation at the open end,
we expect $\lambda_0 = (1.5\pm0.5)$~cm for the experimental range of the aperture diameter
($\Phi \in [0.4,0.7]$~cm).
From the agreement of this last estimate with the experimental value of $\lambda_0$,
we conclude that the offset originates from the radiation of the resonant acoustic wave
at the open end.

\subsection{Damping of the acoustic signal}

\begin{figure}[!h]
\begin{center}
\includegraphics[width=\columnwidth]{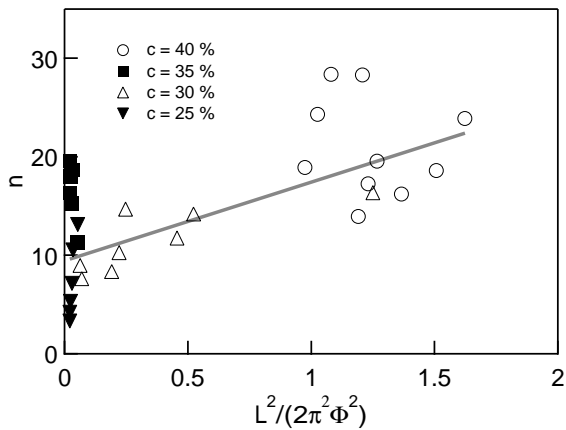}
\end{center}
\caption{\small{
    {\bf Characteristic duration of the sound emission $n$, in number of periods, vs. $L^2/\Phi^2$}.
    The dependence of $n$ on the ratio $L/\Phi$ demonstrates that the damping of the
    acoustic signal is mainly governed by the radiation at the open end
    (The slope of the grey line is 8, thus corresponding to $\xi = \frac{1}{8}$, see text.)}}
\label{fig:damp}
\end{figure}

In order to account for the damping of the acoustic wave, we report
in Fig.~\ref{fig:damp} the typical duration, in number of periods,
$n$, of the acoustic signals. This choice is explained by the difficulty 
in defining precisely the signal duration because of the rather complex envelop 
of the acoustic signal. For an exponential decay over the characteristic
time $\tau$, we would expect $n = n_0 + \frac{\omega\tau}{2\pi}$, so
that the experimental value of $n$ provides an estimate of the
characteristic time $\tau$ to within an offset. The damping of the
acoustic signal might be governed by several physical processes,
among them the partial reflection at the cusp, the viscous
dissipation at the side walls and the diffraction at the open end.
Each of the processes would lead to a different dependence of the
duration on the geometry of the bubble. The partial reflection at
the cusp would lead to constant $\omega \tau$ whereas the viscous
dissipation would lead to $\omega \tau \propto \Phi \sqrt{\omega}$
\cite{Vidal06} and, thus, to a decrease in $n$ for increasing
$L$. On the contrary, one observes experimentally an increase in
$n$ with the bubble length $L$. In order to account for the
radiation, we can further expand the acoustic impedance ${\cal
Z}_{out}$ to the second order in $k \Phi$ and write ${\cal
Z}_{out}/(\rho v) \simeq \zeta j k \Phi + \xi \left(k \Phi\right)^2$
where $\xi$ is a constant. For the radiation of a planar wave at a
flanged ({\it resp.} unflanged) aperture, $\xi = \frac{1}{8}$ ({\it
resp.} $\frac{1}{16}$)  \cite{Rayleigh40, Levine48}. As already
mentioned, we expect the curvature of the wavefront to slightly
alter the value of the impedance to within a term of the order of
$\left({\phi}/{L}\right)^2$ but we do expect $\xi$ to remain of the
same order. The solution of the equation \ref{propagation}, taking 
into account the boundary condition at the open end, then leads to 
$\omega \tau = \frac{1}{\xi} \frac{L^2}{\pi\Phi^2}$ and
thus to $n = n_0 + \frac{1}{\xi} \frac{L^2}{2\pi^2\Phi^2}$. Our
measurements are not accurate enough to make it possible to
determine the experimental slope $\frac{1}{\xi}$ but we point out
that they are compatible with the value ${\xi} = \frac{1}{8}$
(Fig.~\ref{fig:damp}). We can thus conclude that the damping of the
acoustic signal is mainly governed by the radiation at the open end
of the bubble.

\subsection{Acoustic energy}

At this point, it is particularly interesting to focus on the energy associated with the
acoustic emission.
From the pressure signal, $P(t)$, provided by the microphone, one can estimate the total amount
of energy released in the acoustic signal at the fundamental frequency, $E_a$.
Assuming that the acoustic wave outside the bubble is almost spherical, centered at the 
bubble aperture, in the half-space above the free surface, we can write:
\begin{equation}
E_a \simeq \frac{2 \pi d^2}{\rho v} \int_{t=0}^{\infty} P(t)^2 dt
\end{equation}
where, we remind, $d$ stands for the distance from the microphone to the bubble aperture.
As $E_a$ is expected to depend on the volume $V$ and of the initial overpressure $\delta P$
of the air inside the bubble before the bursting, let us now consider an estimate of the
total amount of released energy, $E_T$, assuming a rapid expansion of air:
\begin{equation}
E_T = \frac{1}{2}\frac{V\delta P^2}{\rho v^2}.
\end{equation}
Experimentally, the volume $V$ is obtained from the injected volume
of air but the overpressure $\delta P$ is rather difficult to estimate. However, we
measured the surface tension $\gamma = (25\pm5)$~mN/m of the
gel-water interface for all concentration $c$ and we can estimate
the tension of the thin film at the top, previous to the bursting
event, to be about $2\gamma$. Measuring the radius of curvature of
the bubble head, $\cal R$, from the image, we can estimate $\delta P
\simeq 4 \gamma/{\cal R}$. 
We point out that the corresponding value of $E_T$ must be considered 
with caution (Laplace's law might not be valid for bubbles bursting 
without reaching equilibrium at the free surface). However, it makes possible
to estimate - and therefore, to further discuss - the energy of the acoustic emission.

\begin{figure}[!h]
\begin{center}
\includegraphics[width=\columnwidth]{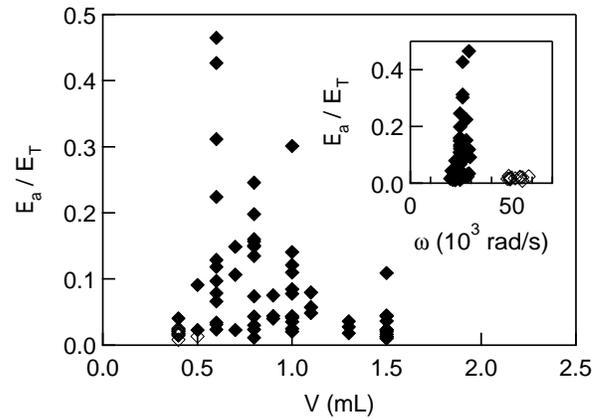}
\end{center}
\caption{\small{{\bf Ratio $E_a/E_T$ vs. volume $V$.}
    At a given volume $V$, the acoustic energy $E_a$ does not account for the amount of energy released
    by the bursting bubble.
    {Inset: Ratio $E_a/E_T$ vs. frequency $\omega$}: The experimental results clearly demonstrate that the
    bubbles can be separated in two categories: small bubbles that sit at the free surface before bursting
    (open symbols) and larger ones that cross the interface without stopping (full symbols).}}
\label{fig:energy}
\end{figure}

We report first $E_a/E_T$ as a function of the bubble volume $V$
(Fig.~\ref{fig:energy}). One clearly observes that the data points
are widely distributed, which indicates that the acoustic energy
$E_a$ does not image the total amount of energy released by the
bursting event, $E_T$. However, the envelop of the data points
indicates that, at intermediate volume, in some cases, a large part
of $E_T$ is transferred to the acoustic mode, which explains why some
of the bubbles are so loud whereas others are barely audible.
As already discussed in Ref.~\cite{Vidal06}, the transfer of the
energy intially loaded in the bubble to the acoustic modes is mainly
governed by the characteristic time associated with the rupture of the bubble head:
The opening of the cavity is efficient in exciting the inner
resonant modes only if rapid enough.
For the smallest volumes, the bubble rises and stops at the free surface. The film that
closes the cavity bursts after drainage.
Even small, the characteristic time associated with the cavity opening is long compared
to the acoustic period because the bubble body is very short.
As a consequence, $E_a/E_T$ is relatively small.
In the opposite limit, the bubbles are large and cross the interface without stopping.
The bursting results from the breaking of the bubble head which is torn apart due the
bubble dynamics.
In this case, the film is not as thin as one would obtain as the result of the drainage
and the characteristic opening-time is long compared to the period associated with the
resonant modes, even if the bubble body is long.
Again, the ratio $E_a/E_T$ is relatively small.
The optimal conditions are reached when the volume of the bubble is such that the bubble
crosses the interface slowly enough for the film to have time to thin but rapidly enough
for the bubble tail not to disappear (this leads to a significant
increase of the resonant frequency).
This latter conclusion is supported by the data reported in Fig.~\ref{fig:energy}, inset. One
observes, reporting $E_a/E_T$ as a function of the resonant
frequency $\omega$, that the data can be separated in two groups:
bubbles that sit at the interface before bursting and bubbles that
cross the interface without stopping.
We clearly note that $E_a/E_T$ is maximum for bubbles crossing dynamically the free surface.

From these remarks on the acoustic energy, we would like to point
out that the acoustic energy in the fundamental mode drastically
depends on the dynamics of the thin film rupture and that, as a
consequence, measuring the acoustic energy is not enough for
obtaining a good estimate of the total energy release.

\section{Conclusion}

Motivated both by the will to understand the physical processes involved when a bubble
bursts at the surface of a non-newtonian fluid, and the hope to use the acoustics as a tool
to investigate natural systems such as volcanoes the relationship between the acoustic
wave and the rheological properties of lava, we have investigated the bursting of bubbles
at the free surface of a gel solution, diluted at different concentration.

We have shown that, at large-enough gel-concentration, the bubble, which is elongated,
acts as a motionless resonator and, thus, exhibits a well-defined acoustic-frequency
at bursting. The amplitude of the acoustic wave emitted
at bursting depends on various parameters: gel concentration, volume of the bubble, and
film rupture time. In spite of the observed clear transition between a static regime, where
the small-volume bubble remains trapped at the surface, and a dynamical regime, where
the high bubble-rising velocity makes the bubble go through the surface and burst directly,
the problem remains rich and complex, in particular, due to the unpredictable film rupture
time, which directly influences the amplitude and energy of the acoustic signal.

From a practical point of view, we can raise the following question:
What pieces of information can we infer from acoustic measurements, if they are
the only available data? From our study, we can conclude that the
frequency of the signal gives a direct access to the bubble length.
However, any attempt to interpret the amplitude and energy of the
acoustic signal would surely lead to strong misinterpretation.
Indeed, we have shown that the same experimental conditions (gel
concentration, bubble volume) can lead to completely different
acoustic-signals at bursting, due to the high sensitivity to the film bursting dynamics.

\vspace{0.3cm}

\begin{center}
{\bf ACKNOWLEDGEMENTS}
\end{center}
The collaborative research was supported by CNRS/CONICYT project \#18640 
and Conicyt-Chile FONDAP project \#11980002.





\label{lastpage}


\begin{thebibliography}{99}

\bibitem{parfitt04}
E.~A. Parfitt, J. Volcanol. Geotherm. Res. {\bf 134}, 77 (2004).

\bibitem{gonnermann03} 
H.~M. Gonnermann and M. Manga, Annu. Rev. Fluid Mech. {\bf 39}, 321 (2007).

\bibitem{weaire99} 
D. Weaire and S. Hutzler, {\it The Physics of Foams} (Oxford University Press, New York, 1999).

\bibitem{prosperetti93} 
A. Prosperetti and H.~N. Oguz, Annu. Rev. Fluid Mech. {\bf 25}, 577 (1993).

\bibitem{leighton94} 
T.~G. Leighton, {\it The Acoustic Bubble} (Academic Press, San Diego, 1994).

\bibitem{muller99} 
W. M\"{u}ller and J.-M. di~Meglio, J. Phys.: Condens. Matter {\bf 11}, L209 (1999).

\bibitem{vandewalle01}
N. Vandewalle, J.~F. Lentz, S. Dorbolo, and F. Brisbois, Phys. Rev. Lett. {\bf 86}, 179 (2001).

\bibitem{vandewalle02} 
N. Vandewalle and  J.~F. Lentz, Phys. Rev. E {\bf 64}, 021507 (2001).

\bibitem{vandewalle03} 
N. Vandewalle, H. Caps, and S. Dorbolo, Physica A {\bf 314}, 320 (2002).

\bibitem{ding07} 
J. Ding, F.~W. Tsaur, A. Lips, and A. Akay, Phys. Rev. E. {\bf 75}, 041601 (2007).

\bibitem{muller07} 
F. M\"{u}ller, U. Kornek, and R. Stannarius, Phys. Rev. E. {\bf 75}, 065302(R) (2007).

\bibitem{Johnson03} 
J.~B. Johnson, R.~C. Aster, M.~C. Ruiz, S.~D. Malone, P.~J. McChesney, J.~M. Lees 
and P.~R. Kyle, J. Volcanol. Geotherm. Res. {\bf 121}, 115 (2003).

\bibitem{jaupart} C. Jaupart and S. Vergniolle, Nature {\bf 331}, 58 (1988);
C. Jaupart and S. Vergniolle, J. Fluid Mech. {\bf 203}, 347 (1989). 

\bibitem{Ripepe01}
M. Ripepe, S. Ciliberto and M. Della Schiava, J. Geophys. Res. {\bf 106}, 8713 (2001).

\bibitem{Webb90}
S.~L. Webb and D.~B. Dingwell, J. Geophys. Res. {\bf 95}, 15695 (1990).

\bibitem{Vergniolle96a}
S. Vergniolle and G. Brandeis, J. Geophys. Res. {\bf 101}, 20433 (1996).

\bibitem{Francalanci89}
L. Francalanci, P. Manetti and A. Pecerillo, Bull. Volcanol. {\bf 51}, 335 (1989).

\bibitem{Ichihara05}
M. Ichihara, T. Yanagisawa., Y. Yamagishi, H. Ichikawa, and K. Kurita, Japan Earth 
and Planetary Science Joint Meeting, Abstract (CD-ROM) A111-P004  (2005).

\bibitem{Vidal06}
V. Vidal, J.-C. G\'eminard, T. Divoux and F. Melo, Eur. Phys. J. B {\bf 54}, 321 (2006).

\bibitem{Bird87}
R.~B. Bird, R.~C. Armstrong and O. Hassager, {\it Dynamics of Polymeric Liquids}, 
Vol. I and II (Wiley, New York, 1987).

\bibitem{Podgorski02a}
T. Podgorski and A. Belmonte, J. Fluid Mech. {\bf 460}, 337 (2002).

\bibitem{Hassager79}
O. Hassager, Nature {\bf 279}, 402 (1979).

\bibitem{Liu95}
Y. Liu, T. Liao and D.~D. Joseph, J. Fluid Mech. {\bf 304}, 321 (1995).

\bibitem{Chhabra}
R.~P. Chhabra, {\it Bubbles, drops, and Particles in Non Newtonian fluids}, 2nd edition 
(Chemical industries series \#113, Taylor \& Francis, 2007). 

\bibitem{Handzy04}
N.~Z. Handzy and A. Belmonte, Phys. Rev. Lett. {\bf 92}, 124501 (2004).

\bibitem{Jayaraman03}
A. Jayaraman and A. Belmonte, Phys. Rev. E {\bf 67}, 065301 (2003).

\bibitem{Belmonte00}
A. Belmonte, Rheol. Acta {\bf 39}, 554 (2000).

\bibitem{Dalkoni99}
K. Dalkoni-Veress, B.~G. Nickel, C. Roth \emph{et al.}, Phys. Rev. E {\bf 59}, 2153 (1999).

\bibitem{Debregeas95}
G. Debr\'egeas, P. Martin and F. Brochard-Wyart, Phys. Rev. Lett. {\bf 75}, 3886 (1995).

\bibitem{Debregeas98}
G. Debr\'egeas, P.-G. de Gennes and F. Brochard-Wyart, Science {\bf 279}, 1704 (1998).

\bibitem{Dutcher00}
J.~R. Dutcher, K. Dalkoni-Veress, B.~G. Nickel \emph{et al.}, Macromol. Symp. 
{\bf 159}, 143 (2000).

\bibitem{Warszynski96}
P. Warszy\'nski, B. Jachimska and K. Malysa, Colloids Surfaces A: Physicochem. Eng. Aspects 
{\bf 108}, 321 (1996);  B. Jachimska, P. Warszy\'nski and K. Malysa, Colloids Surfaces A: Physicochem. Eng. Aspects 
{\bf 143}, 429 (1998).

\bibitem{Angarska01}
J.~K. Angarska and E.~D. Manev, Colloids Surfaces A: Physicochem. Eng. Aspects {\bf 190}, 117 (2001).

\bibitem{Levine48} 
H. Levine and J. Schwinger, Phys. Rev. {\bf 73}, 383 (1948).

\bibitem{Landau87}
L.~D. Landau and E.~M. Lifshitz, \emph{Course in Theoretical Physics: Fluid Mechanics},
vol.6, 536 pp. (Pergamon, Tarrytown, N.Y., 1987).

\bibitem{Pierce89}
A. Pierce, {\it Acoustics - An introduction to its physical principles and applications} 
(ASA, New-York, 1989).

\bibitem{Rayleigh40}
Lord Rayleigh, {\it Theory of Sound} (Macmillan, 1940).

\end{thebibliography}
\end{document}